\documentclass[12pt]{article}
\usepackage[utf8]{inputenc}
\usepackage{amsmath, amssymb}
\usepackage{graphicx}
\usepackage{booktabs}
\usepackage{array}
\usepackage{caption}
\usepackage{subcaption}
\usepackage{geometry}
\usepackage{authblk}
\usepackage{natbib}
\usepackage{hyperref}

\hypersetup{
    colorlinks=true,
    linkcolor=blue,
    citecolor=blue,
    urlcolor=blue
}

\title{Forecasting Economically Significant Bitcoin Moves: A Multi-Scale TCN with Profit-Optimized Thresholds}

\author[1]{Parsa Yousefnezhad}
\author[1]{Gholamreza Mansourfar\thanks{Corresponding author: g.mansourfar@urmia.ac.ir}}
\author[2]{Mohammadreza Feizi Derakhshi}
\affil[1]{Department of Accounting, Faculty of Economics and Management, Urmia University, Urmia, Iran}
\affil[2]{Department of Computer Engineering, Faculty of Electrical and Computer Engineering, University of Tabriz, Tabriz, Iran}

\date{}

\begin{document}

\maketitle

\begin{abstract}
Bitcoin's future fluctuations are a substantial concern for investments and risk management. Investors and financial institutions require accurate forecasts of these price movements to hedge and optimize portfolios. This study aims to answer the question of whether Bitcoin's price will rise beyond 5\% within the next 7 days by utilizing on-chain, market, and sentiment data from February 2018 to December 2025. The proposed model consists of a multi-scale temporal convolutional network with InceptionTCN blocks, CNN channel attention, adaptive average pooling, and a pairwise ranking loss. Dilated convolutions with bottleneck and fusion layers are employed to efficiently capture features over horizons from 1 to 4 days. Given the class imbalance in the dataset, AUC is used instead of accuracy and other classification metrics to reflect the model's performance better. Subsequently, a profit-optimized decision threshold is also applied to align model selection with financial objectives. The proposed model is compared with 5 other baselines: ImprovedTCN\_GRU, LSTM, TCN, XGBoost, and Random Forest. Results indicate that the proposed model achieved an AUC of 0.6316 and a profit of 1.703, outperforming all baseline models. Using a novel deep learning model would assist investors in making better financial decisions.
\end{abstract}


\section{Introduction}

The introduction of Bitcoin by Satoshi Nakamoto in 2008 revolutionized the finance industry, with the market value of the virtual asset being more than \$1.2 trillion as of June 2026 \citep{nakamoto2008bitcoin}. Cryptocurrencies appeal to individual investors and traders, and academics alike \citep{nofer2017blockchain}. They are built on blockchain technology and are open, permanent, and resistant to central authority.

The financial world has been revolutionized with the introduction of digital currency, particularly a decentralized virtual currency, Bitcoin, which is highly volatile, liquid, and a secure and reliable asset class \citep{jagannath2021self}. Unlike other currencies or stocks, the price of Bitcoin is not easily predictable, and its movements can be influenced by macroeconomic indicators, on-chain activity, speculation, the hash rate, and the difficulty setting \citep{jang2018empirical, jagannath2021self, kristoufek2023bitcoin}.

These differences provide a tremendous opportunity for quantitative trading but are a significant obstacle in prediction modeling and risk management. Such changes can come with significant challenges for the participants to adapt to, and it is crucial to have accurate and stable directional forecasting models to make decisions, optimize portfolios, and manage risks \citep{alhnaity2020new, chen2023analysis}. Hence, financial institutions and individuals need to have accurate forecasting models. At short horizons, Bitcoin returns have very high kurtosis (fat tails) and frequent extreme moves, which introduce various economic opportunities to gain profit \citep{kose2024bitcoin}. Many scientific publications analyze the direction and trend of the price that can be predicted \citep{kuznetsov2025machine}.

However, focusing solely on the direction is not adequate. Forecast models should account for the magnitude of returns too and distinguish between 1\% and 10\% changes, which is vital in volatile markets such as the cryptocurrency market \citep{catania2019forecasting}. Direction forecasting models ignore real trading profitability, leading to an unrealistic modeling of asset movements \citep{critien2022bitcoin}. Moreover, many studies attempt to optimize accuracy, since Bitcoin data usually faces class imbalance; using accuracy and other classification metrics is misleading \citep{loginova2024forecasting}. To have a better comparison under the class imbalance situation, the AUC (Area Under the Receiver Operating Characteristic Curve) metric is used to compare the model's performance \citep{kamalov2020forecasting}.

The primary objective of this research is to determine whether a multi-scale ImprovedTCN, incorporating attention mechanisms and pairwise ranking loss, can outperform recurrent, convolutional, and tree-based models in terms of AUC evaluation and profit-based decision thresholds.

This paper aims to forecast whether Bitcoin will exceed 5\% within 7 days. This 5\% threshold is chosen not only to capture significant and actionable moves by filtering out noisy moves for day traders, but also to cover transaction costs confidently. A profit function is also applied to evaluate the model's profitability and ensure that captured signals cover transaction costs confidently \citep{karzanov2024headline}. This profit function aligns models' performance with the financial objectives, which is vital for traders and investors to make better financial decisions.

The proposed model combines multi-scale inception blocks that process the input sequence through parallel convolutional branches with different dilated convolutions. This enables the model to simultaneously capture short-term, medium-term, and longer-term temporal dependencies \citep{thill2021temporal}. A bottleneck layer is added before the parallel branches to reduce the channel dimension, followed by a fusion layer to merge the outputs from all dilated branches. The bottleneck-fusion design is adopted to enhance parameter efficiency and encourage feature mixing, which is significant when working with high-dimensional on-chain data \citep{spadon2024multi}. After the multi-scale TCN captures local and multi-frequency dependencies, a CNN channel attention + Adaptive Average Pooling classifier is used to attend to informative features and time steps \citep{metawa2023fraud}.

The proposed ImprovedTCN model is compared against five other competitive models: a hybrid ImprovedTCN\_GRU, a standard LSTM, a plain TCN, XGBoost, and Random Forest. These baseline models cross recurrent, convolutional, and tree-based approaches. This comparison allows us to assess the proposed model's advantages in terms of AUC and profit-based factor.

A rigorous walk-forward backtest with five splits is applied to retrain and recalibrate the model on expanding training windows. This design addresses the look-ahead bias and includes non-stationarity in financial data \citep{bieganowski2025supervised}. A comprehensive suite of on-chain features, including hashrate, realized capitalization, NUPL, MVRV, spent output profit ratios, and fear \& greed index, is also integrated to capture fundamental network activity and investor behaviors \citep{bouteska2024cryptocurrency}.

The rest of this paper is organized as follows. The related works regarding Bitcoin price prediction, deep learning in time series, and profit evaluation are shown in Section 2. The methodology is explained in Section 3, featuring the ImprovedTCN architecture, baseline models, and pairwise ranking loss. Section 4 presents the empirical results and compares all six models. Section 5 discusses the implications of our findings, why ImprovedTCN outperforms the baselines, and the limitations. Section 6 summarizes the contributions, offers suggestions for practice, and projects future research.

\section{Literature review}

This section explains four interrelated fields of related literature: Bitcoin price prediction approaches, imbalanced data challenges, profit-based assessment in financial forecasting, and multi-scale temporal architectures, on-chain and sentiment analysis. This review contains results from econometrics, machine learning, and deep learning models, and puts them in the perspective of the contribution of the proposed ImprovedTCN framework and the gaps that motivate this research.

\subsection{Bitcoin Price Forecasting Methodologies}

The early models of predicting bitcoin price were the classical econometric models, such as Autoregressive Integrated Moving Average (ARIMA) and Generalized Autoregressive Conditional Heteroskedasticity (GARCH). For modelling linear dependency and volatility clustering, the following statistical models are used. Non-linear, structural breaks, and extreme events are crucial in the cryptocurrency markets \citep{sabbour2025comparison}. To extract any non-linear dependency in machine learning, models such as SVM, RF, and XGBoost have been developed, and these models are more effective for time series data \citep{rao2025new}.

Typical machine learning models, however, struggle to capture the complex multi-scale temporal correlations, volatility, and non-linear relationships that are intrinsic to the cryptocurrency market \citep{rafi2024cryptocurrency}. Deep learning methods have been found to be highly promising in capturing long-range patterns, which traditional machine learning models face limitations in doing so \citep{alizadegan2024forecasting, wang2025meta}. \citet{fischer2018deep} predict out-of-sample daily returns of the S\&P 500 using LSTM and achieve a 0.46 percent daily return, better than other machine learning methods such as Random Forest, Logistic Regression, and Deep Neural Network. Later, LSTM, GRU, and Bi-LSTM were compared on Bitcoin, Ethereum, and Litecoin, which Bi-LSTM achieved better performance, with a MAPE of 0.036 for Bitcoin \citep{seabe2023forecasting}.

Recurrent architectures, however, do have certain limitations in computation, sequential processing, and modelling long-range dependencies \citep{dave2025multi}. Recently, Temporal Convolutional Networks (TCNs) have been used because of their parallelizable architectures, stable gradients, and dilated convolutions that are used to capture multi-scale patterns \citep{wu2025tcn}. As shown by \citet{bai2018empirical}, TCNs can be trained to have a long memory, stable gradients, and parallel computation that is not possible with recurrent architectures. Another study by \citet{chen2020probabilistic} shows that causal convolutions as well as residual blocks can successfully learn intricate patterns.

\subsection{Handling Class Imbalance and Profit-Oriented Evaluation}

Financial time series data suffer from severe imbalance in the data, and there are many small fluctuations in price. Consequently, models learn to forecast the majority class with high accuracy while missing economically important but rare price movements \citep{basu2022novel}. \citet{verbraken2013novel} demonstrated that aligning model selection with respect to economic objectives is important to evaluate the financial performance of the models. Furthermore, \citet{jensen2026machine} argued that forecasting models must be evaluated on their risk-return trade-off by introducing the concept of the ``implementable efficient frontier''. Other studies such as \citet{omole2024deep} and \citet{sebastiao2021forecasting} implemented backtesting strategies. However, systematic profit-based decision threshold optimization remains absent.

\subsection{Multi-Scale and Attention Mechanisms in Time Series}

TCNs use dilated convolutions to extract long-range dependency with only a linear number of parameters. This allows an exponentially expanding receptive field, while at the same time providing efficient training and stable gradients from recurrent architectures \citep{chang2017dilated, chen2020probabilistic}. The inception-style architectures initially developed for computer vision (CV) are based on multiple parallel branches of small convolutional kernels for capturing multi-scale information in parallel, and have been adapted to time-series forecasting, where multi-scale information refers to patterns learned across different time horizons \citep{benidis2022deep, chen2023multi}.

Attention mechanisms have revolutionized the area of sequence modeling, making models that selectively attend to the time steps and feature channels that are important. These methods have been adopted by forecasting methods more and more. However, only a few studies have explored multi-scale convolutions, channel attention in a common framework to forecast cryptocurrencies in particular.

The proposed ImprovedTCN addresses this by using multi-scale convolutions with dilations of 1, 2, and 4, a bottleneck fusion architecture for parameter efficiency, and CNN channel-level attention for adaptive feature selection \citep{aryal2020comparative, han2024capacity}. This systematic combination of well-known methods, namely multi-scale TCN, channel attention, and pairwise ranking loss, is specific to the Bitcoin forecasting problem. The individual parts are known, but the mixture and use in this context is new.

\subsection{On-Chain data and Sentiment Analysis}

The blockchain's transaction records provide cryptocurrencies with unique data modalities. \citet{kukacka2023fundamental} showed the impact of on-chain activities on Bitcoin prices, while \citet{kim2022deep} presented a deep learning model specifically based on the on-chain data, showing that it outperforms on-chain price and volume-only models in predicting price changes. At the same time, sentiment gathered from social media has been beneficial. \citet{valencia2019price} demonstrated that sentiment from platforms such as Twitter can improve prediction accuracy, while \citet{huynh2021does} and \citet{critien2022bitcoin} demonstrated that sentiment from Twitter can improve the accuracy of a sentiment classification task. However, the best integration of those three different modalities -- on-chain, market and sentiment -- is not systematically explored within a single coherent deep learning architecture, and the ability to model all the drivers of a cryptocurrency's prices is still incomplete.

\subsection{Gap Identification and Research Contributions}

Existing approaches largely neglect the benefits of multi-scale temporal modelling. One problem with standard TCNs is that they use fixed dilation patterns instead of the simultaneous occurrence of short-term price increases, medium-term trend reversals, and long-term market cycles, which characterize the behavior of cryptocurrencies. To overcome this, the InceptionTCN architecture with multiple parallel dilations (1, 2, 4) proposed in our work offers a receptive field for multiple temporal resolutions without an exponential growth in parameters. In addition, existing models tend to process all features equally across all channels, ignoring the market regime-specific differential predictive power of on-chain indicators compared to sentiment indicators. CNN channel attention: the model dynamically assigns weights to the channels depending on the context, allowing it to prioritize the most informative signal. Likewise, some architectures rely on basic mean or max-pooling in the temporal dimension to lose out on the fine-grained importance patterns in time. We use a pairwise ranking loss function that directly optimizes the AUC, which is the most relevant metric in imbalanced binary classification, by favoring higher scores from positive examples over negative examples. The multi-scale dilated convolutions, coupled with the adaptive channel attention, learnable temporal weighting, and AUC-optimizing pairwise ranking loss, are a novel synthesis directly aimed at the identified shortcomings of the existing literature. This framework combines a comprehensive suite of on-chain and market features over a 7-year timeframe (2018--2025), along with strict walk-forward validation with a profit-optimized threshold, providing a strong, economically sound solution for cryptocurrency trading.

\section{Methodology}

\subsection{Data and sample}

In this study, an extensive dataset is built from February 2018 to December 2025, consisting of three types of predictive signals: market microstructure, on-chain fundamentals, and sentiment indicators (Fear \& Greed Index, Google Trends). This multi-modal dataset enables the model to account for the various factors influencing Bitcoin price fluctuations, as mentioned in the literature review.

The descriptive statistics for the market data are reported in \autoref{tab:market_stats}. The average closing price was \$37,456, ranging from \$3,212 to \$124,625. Daily volume is highly skewed (3.531) and highly leptokurtic (16.367), suggesting that extremely large liquidity events are rare. Bitcoin returns are fat-tailed, asymmetric, and negatively skewed ($-1.128$), with a near-zero mean of the daily log returns (0.000078) and very high kurtosis (18.743).

\begin{table}[htbp]
\centering
\caption{Descriptive statistics of daily Bitcoin market data}
\label{tab:market_stats}
\begin{tabular}{lrrrrrrr}
\toprule
Variable & Count & Mean & Std & Min & Max & Skewness & Kurtosis \\
\midrule
Open   & 2888 & 37048.9 & 32255.9 & 3211.7 & 124658.5 & 0.978 & -0.099 \\
High   & 2888 & 38226.1 & 32784.4 & 3276.5 & 126199.6 & 0.961 & -0.141 \\
Low    & 2888 & 36597.6 & 31689.3 & 3156.3 & 123084.0 & 0.995 & -0.053 \\
Close  & 2888 & 37456.8 & 32265.6 & 3211.7 & 124625.5 & 0.976 & -0.104 \\
Volume & 2888 & 65695.7 & 77887.4 & 1521.5 & 760705.4 & 3.531 & 16.367 \\
Returns & 2888 & 0.000078 & 0.0343 & -0.5026 & 0.1784 & -1.128 & 18.743 \\
\bottomrule
\end{tabular}
\end{table}

An Augmented Dickey-Fuller (ADF) test was performed on the 7-day forward return series to test for stationarity. The test statistic was $-7.937$, and p-value $< 0.001$, which is below the 5\%. Therefore, we strongly reject the null hypothesis of a unit root, confirming that the series is stationary. This property justifies the use of the raw returns series for modeling without requiring differencing. Results of the ADF test are presented in \autoref{tab:adf}.

\begin{table}[htbp]
\centering
\caption{ADF test results}
\label{tab:adf}
\begin{tabular}{lcccc}
\toprule
Variable & Test Statistic & P-Value & Observations & Conclusion \\
\midrule
7-Day Forward Return & -7.937 & 0.001 & 2852 & Stationary \\
\bottomrule
\end{tabular}
\end{table}

\autoref{tab:onchain_sentiment} summarizes the on-chain and sentiment features. On-chain metrics are highly persistent and positively skewed, such as hashrate, with a mean of 319 million TH/s and a skewness of 1.13. The realized profit/loss ratio has a very highly leptokurtic (kurtosis equal to 71.3) distribution. Sentiment indicators range around neutral (Google Trends mean 45.1, Fear \& Greed mean 47.2).

\begin{table}[htbp]
\centering
\caption{Selected summary statistics for daily on chain and sentiment features}
\label{tab:onchain_sentiment}
\begin{tabular}{lrrrr}
\toprule
Variable & Mean & Std & Skewness & Kurtosis \\
\midrule
Hashrate (TH/s)          & 318983127.2 & 302074022.4 & 1.1300 & 0.1391 \\
realized\_price (USD)    & 19808.1     & 14320.3979  & 0.8467 & -0.0060 \\
nupl\_7dma               & 0.3597      & 0.2308      & -1.0417 & 0.5432 \\
realized\_profit\_loss\_ratio & 5.0620 & 9.9075 & 6.4626 & 71.3060 \\
fear\_greed              & 47.1485     & 21.7135     & 0.1628 & -1.0371 \\
google\_trends           & 45.0633     & 23.4065     & 0.9651 & 0.0509 \\
\bottomrule
\end{tabular}
\end{table}

All of the selected features have been proven to signal changes in the cryptocurrency market. A full list of utilized features is presented in \autoref{tab:feature_list}. On-chain metrics such as hashrate, MVRV, and NUPL offer insights into the network fundamentals that frequently precede price breakouts. For example, surges in mining activity and positive valuation trends often indicate accumulation periods. Sentiment indicators capture speculative behavior and are considered to have a more amplified effect on short-term returns, in this case over the last 7 days, which is where retail sentiment has been found to have the strongest influence on price momentum. Market microstructure variables reflect the liquidity conditions required to support a 5\% move; low liquidity can prevent prices from reaching the threshold even when fundamentals indicate upside potential.

\begin{table}[htbp]
\centering
\caption{List of all features}
\label{tab:feature_list}
\begin{tabular}{p{0.45\linewidth}}
\toprule
hashrate \\
hashprice \\
google\_trends \\
address\_10k\_1k \\
true\_market\_mean \\
realized\_price \\
fng\_index \\
nvts\_bg \\
addresses\_active \\
nupl\_7dma \\
nrpl \\
delta\_cap \\
realized\_loss \\
realized\_profit \\
realized\_profit\_loss\_ratio \\
market\_cap \\
investor\_cap \\
thermo\_cap \\
realized\_cap\_capitalization \\
thermocap\_realizedcap\_ratio \\
realized\_cap\_capitalization\_pct \\
coin\_10k \\
coin\_10k\_pct \\
coin\_10k\_1k \\
coin\_10k\_1k\_pct \\
choppiness\_7 \\
choppiness\_30 \\
choppiness\_5w \\
choppiness \\
open \\
high \\
low \\
total\_vol\_proxy \\
\bottomrule
\end{tabular}
\end{table}

The Pearson correlation matrix for the important variables is provided in \autoref{tab:correlation}. The hashrate and realized capitalization are highly correlated with the market price (0.89 and 0.94, respectively), illustrating the long-term co-movement between network security and valuation. The price shows a weak negative correlation with volume ($-0.24$) and hashrate ($-0.19$), meaning that higher volume and hashrate days tend to be associated with slightly lower price days. The price sentiment indicators correlate moderately with price, as bullish sentiment is associated with rising prices (fear\_greed: 0.31, google\_trends: 0.62).

\begin{table}[htbp]
\centering
\caption{Pearson correlation matrix of selected market, on chain and sentiment variables}
\label{tab:correlation}
\begin{tabular}{lrrrrrr}
\toprule
Variable & close & volume & hashrate & fear\_greed & google\_trends & realized\_cap \\
\midrule
close          & 1.00  & -0.24  & 0.89     & 0.31        & 0.62          & 0.94 \\
volume         & -0.24 & 1.00   & -0.19    & -0.12       & -0.02         & -0.11 \\
hashrate       & 0.89  & -0.19  & 1.00     & 0.21        & 0.62          & 0.94 \\
fear\_greed    & 0.31  & -0.12  & 0.21     & 1.00        & 0.17          & 0.11 \\
google\_trends & 0.62  & -0.02  & 0.62     & 0.17        & 1.00          & 0.61 \\
realized\_cap  & 0.94  & -0.11  & 0.94     & 0.11        & 0.61          & 1.00 \\
\bottomrule
\end{tabular}
\end{table}

\begin{figure}[htbp]
\centering
\includegraphics[width=\linewidth]{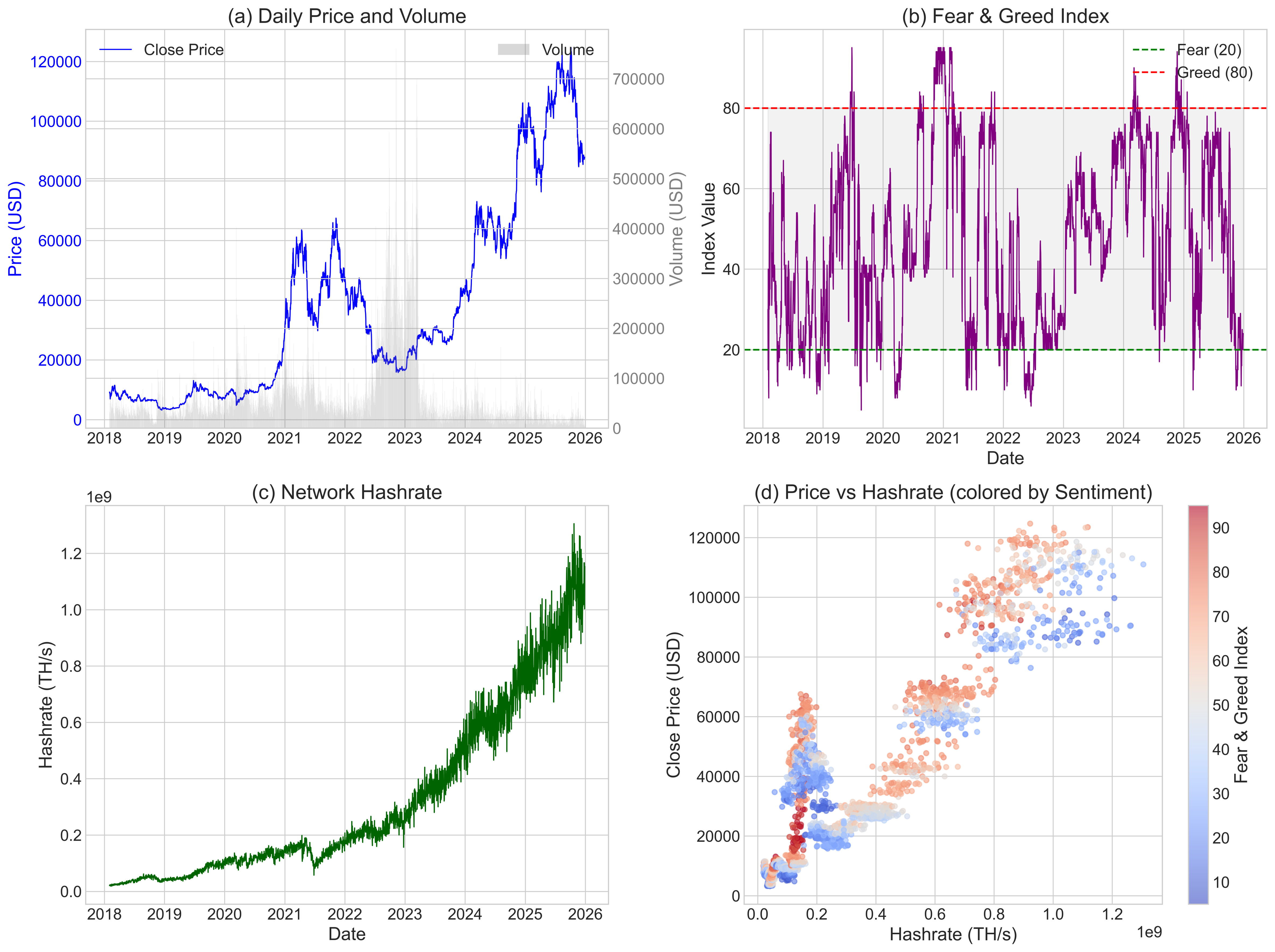}
\caption{Comprehensive visualization of Bitcoin market dynamics. (a) Hourly price evolution overlaid with trading volume. (b) Daily Fear \& Greed Index fluctuations, with thresholds for extreme fear ($<20$, red zone) and extreme greed ($>80$, green zone) marked. (c) Exponential growth of the Bitcoin network hash rate. (d) Scatter plot of Price vs. Hash Rate, colored by sentiment.}
\label{fig:dynamics}
\end{figure}

\autoref{fig:dynamics} shows a complete visualization of the market dynamics of Bitcoin. The daily closing price is plotted overlaid with trading volume (Panel (a)); trading volume surges tend to be seen at significant price inflection points (e.g., March 2020, May 2021, November 2022). Panel (b) plots the Fear \& Greed Index with fear and greed thresholds at 20 and 80, respectively. Extreme sentiment often precedes price reversals. Panel (c) shows that while price drops caused hash rate declines, the network hash rate continued to grow overall despite these price decreases. The scatter plot of price vs hash rate with ``greed'' (orange-red) points dominating the top right quadrant has the highest prices coinciding with the highest security of the network and optimism (highest index).

\begin{figure}[htbp]
\centering
\includegraphics[width=\linewidth]{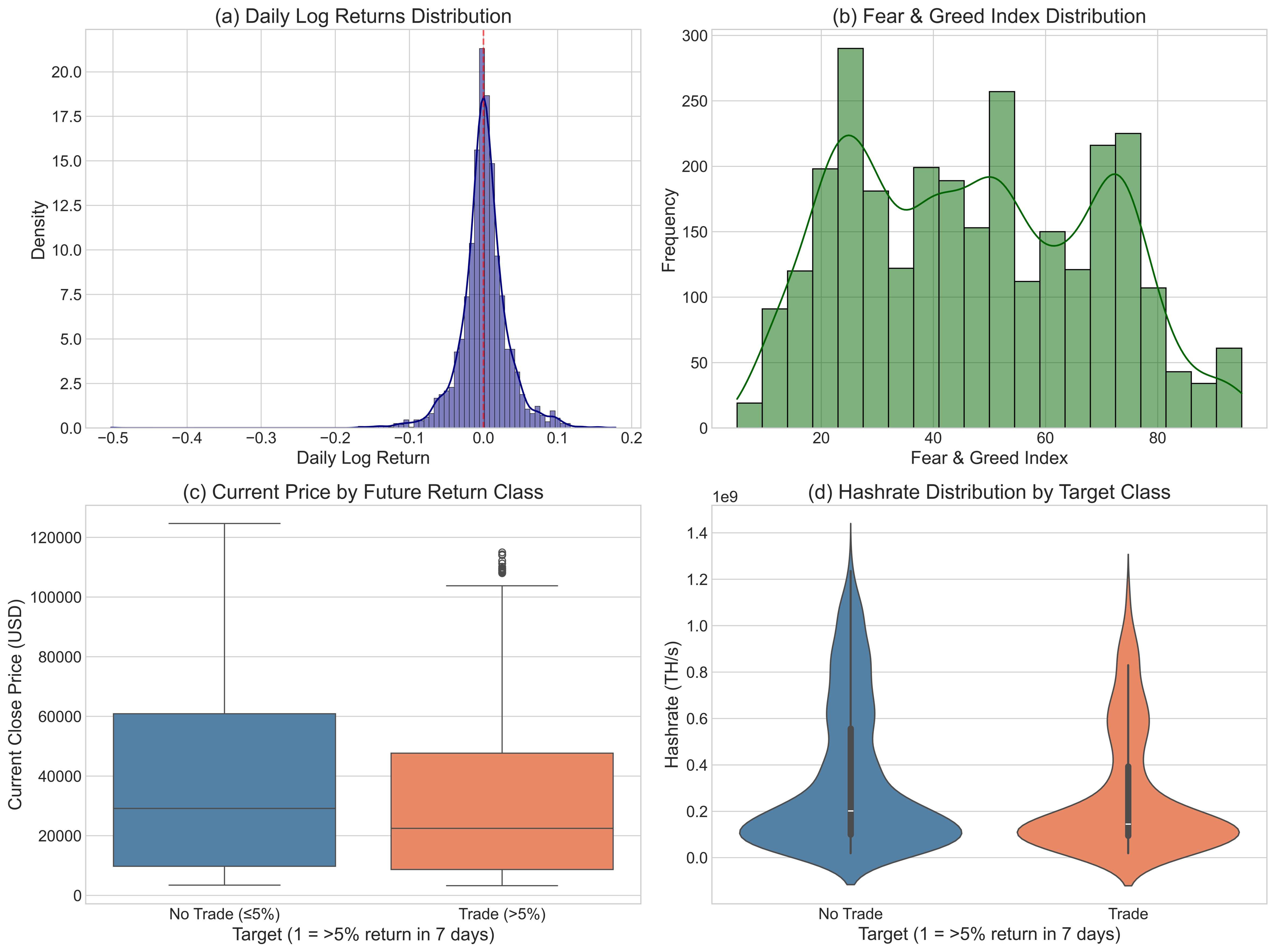}
\caption{Statistical distribution of key features and target class separation. (a) Histogram of hourly returns. (b) Bimodal distribution of the daily Fear \& Greed Index. (c) Box plots of closing prices stratified by target class. (d) Violin plots of hash rate distributions for each target class.}
\label{fig:distributions}
\end{figure}

The distributional properties are analyzed in \autoref{fig:distributions}. Panel (a) presents a histogram of daily log returns, which is leptokurtic with wide tails. The vertical red line shows the 5\% forward return threshold 7 days ahead. The histogram of the Fear \& Greed Index is shown in Panel (b), which is approximately bell-shaped with mass at the extremes. Panel (c) shows boxplots of the current closing price stratified by the future target class (1 if 7-day forward return $>5\%$). The medians are similar, indicating positive targets are across a broad range of price levels. Violin plots of hash rate by target class are shown in Panel (d), and both classes have similar distributions, indicating that significant upward moves are not restricted to certain network regimes.

\begin{figure}[htbp]
\centering
\includegraphics[width=\linewidth]{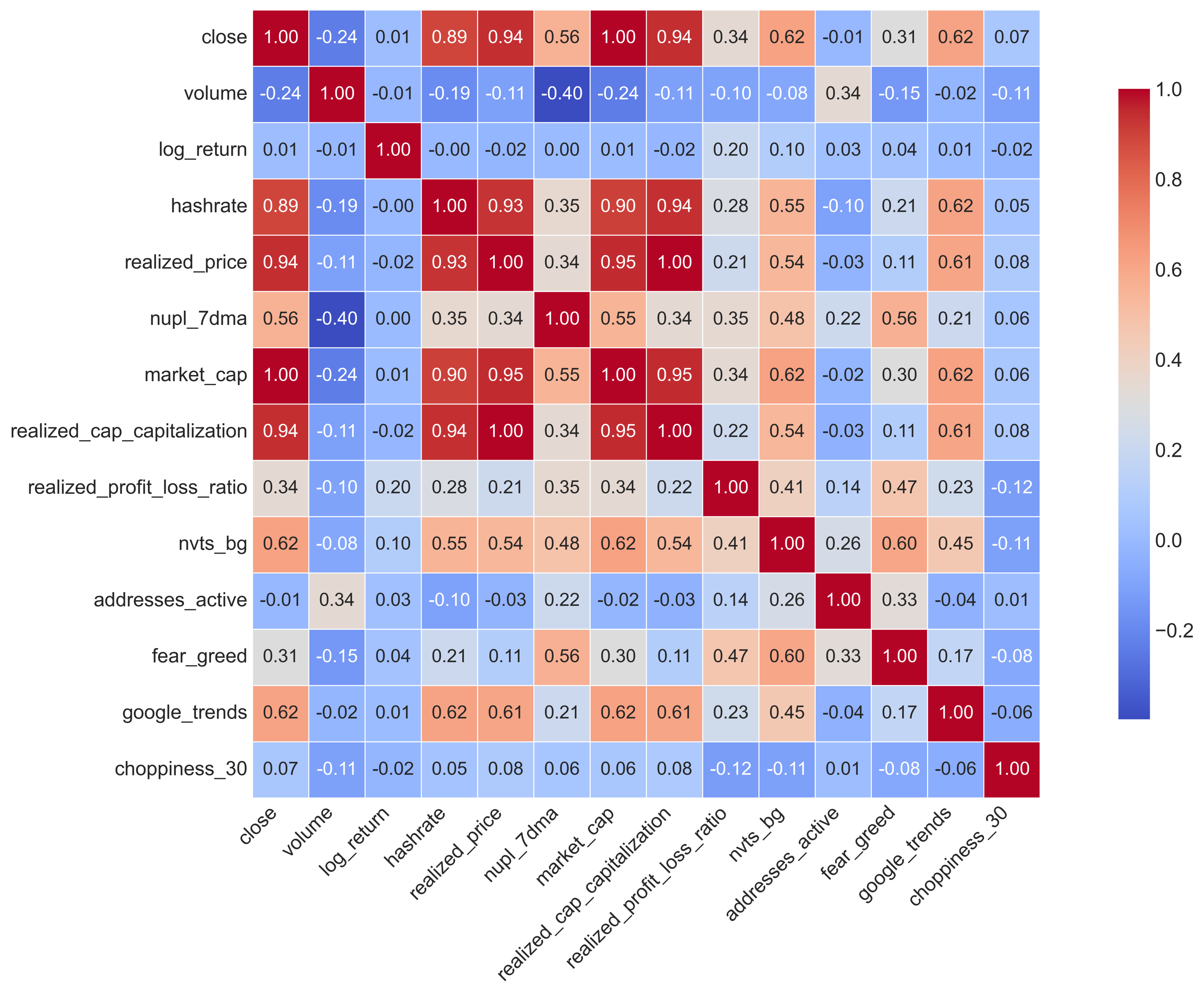}
\caption{Correlation matrix for selected key variables. The strength of the linear relationship is represented by color (red = positive, blue = negative). Hashrate, realized capitalization, and market capitalization exhibit high multicollinearity and strong correlations. In contrast, volume and sentiment indicators show weaker correlations with market capitalization.}
\label{fig:corr_matrix}
\end{figure}

\autoref{fig:corr_matrix} displays the correlation matrix for a subset of key variables. Among this subset, on-chain metrics, like hashrate, realized capitalization, market capitalization, and coin\_10k, exhibit high positive correlations and multicollinearity (red blocks). This is expected because these variables have a similar trend over the longer term. Sentiment indicators (fear\_greed, google\_trends) and volume, on the other hand, exhibit weaker correlations with each other and with the on-chain indicators. The orthogonality is useful because it can be used as complementary signals, which provide non-redundant information to the predictive model.

The target is a binary variable indicating whether the 7-day forward return exceeds 5\%. Formally, $y_{t} = 1$ if $\left( \frac{P_{t + 7} - P_{t}}{P_{t}} > 0.05 \right)$, and 0 otherwise. The 5\% threshold is selected for three economic reasons: (i) the net profitability is guaranteed as the round-trip transaction cost is unlikely to exceed 0.1\% for Bitcoin; (ii) Bitcoin's 7-day volatility is in the range of 8-10\%, meaning that the 5\% move is a reasonable but economically significant signal; and (iii) the positive class proportion (26\%) is large enough to avoid trivial guesswork but small enough that accuracy would be misleading. As a result, AUC and profit-based threshold optimization are chosen (detailed in Section 3.3) as the main evaluation metrics.

\subsection{Preprocessing}

A walk-forward validation scheme is used to test models in a realistic out-of-sample situation, without introducing any look-ahead bias. The full time series (February 2018 - December 2025) is divided into 5 folds of 100 days each using TimeSeriesSplit with n\_splits = 5 and test\_size = 100. For each fold $f$, the training set consists of all observations prior to the $f$-th test window, and the test set is the next 100 days. This expanding-window design mimics a live trading environment where a model is trained only on past data and then evaluated on unseen future data. Importantly, no information in the test set (nor any future observation) is used in the training, hyperparameter tuning, or threshold selection process, thereby avoiding look-ahead bias---as is standard for financial machine learning.

All features are standardized using Z-scores:
\begin{equation}
Z_{i,t} = \frac{x_{i,t} - \mu_{i}^{(\text{train})}}{\sigma_{i}^{(\text{train})}}
\end{equation}
where $\mu_{i}^{(\text{train})}$ and $\sigma_{i}^{(\text{train})}$ are the mean and standard deviation computed exclusively on the training set of each fold. The same scaling parameters are used for the same test set. This per-fold standardization ensures that no leakage from the test distribution is introduced into the training process.

For each time step $t$, an input sequence $X_{t} = [X_{t - 59}, X_{t - 58}, \ldots, X_{t}]$ consists of the last 60 consecutive daily observations of all features. A time frame of 60 days is a sufficient period to detect monthly trends, on-chain cycles, and recurring patterns, but not so long that it becomes impractical to compute and also prohibits the use of information that may be stale by the time the data is analyzed.

\subsection{Model architecture}

The proposed model consists of an efficient multi-scale TCN with InceptionTCNblocks, an integration of CNN channel attention to extract important temporal features across channels and time steps.

\subsubsection{InceptionTCNBlock (Multi-scale TCN core)}

The TCN part consists of three InceptionTCNBlock layers. Each block captures patterns at multiple time horizons while keeping the number of parameters constant.

\begin{itemize}
\item \textbf{Bottleneck:} A $1 \times 1$ convolution reduces the input channel dimension by half. This reduces the computational cost and encourages a more compact representation before the more expensive dilated convolutions.
\item \textbf{Multi-scale dilations:} Three parallel branches with dilations 1, 2, 4 enable the network to capture very local patterns (dilation 1) as well as dependencies that extend to up to $2^{2} = 4$ time steps (dilation 4).
\item \textbf{Fusion layer:} A 1$\times$1 convolution layer fuses the output of all four branches, learning adaptive weights to indicate the importance of each scale.
\item \textbf{Residual connection:} The input is fed into the output of the fusion layer to allow gradient flow and deeper architectures.
\end{itemize}

After three InceptionTCNBlock layers (each producing 64 output channels), a sequence of feature vectors $h_{1}, h_{2}, \ldots, h_{60}$ (one per time step) is produced.

\subsubsection{Improved Classifier}

An improved classifier is also applied to significantly boost the model's capability to focus on the important and relevant features and time steps, which leads to better AUC and profit.

\begin{itemize}
\item A \textbf{Channel Attention} mechanism is implemented to adaptively re-weight the feature channels of the TCN model based on their performance. This specific mechanism enables the model not only to ignore noisy and irrelevant patterns but also to learn which channels are important to emphasize.
\item An \textbf{Adaptive Average Pooling} is utilized in order to aggregate temporal features.
\item The \textbf{dropout layer} is finally applied to help regularize and prevent overfitting.
\end{itemize}

This classifier is utilized to assist the model in adaptively focusing on the most informative time steps and feature channels.

\subsection{Training protocol and loss function}

The proposed model is exclusively trained with a pairwise ranking loss function instead of the standard binary cross-entropy (BCE) loss to directly improve the quality of the ranking of predictions --- a critical requirement for imbalanced financial data. However, Baseline models are trained with the standard and typical BCE loss function. The loss encourages the model to give higher logits to the positive examples (future 7-day return $>5$\%) than the negative ones, with a margin of $m = 1$.
\begin{equation}
\mathcal{L}_{\text{rank}} = \frac{1}{|P||N|}\sum_{i \in P}\sum_{j \in N} \max\left(0, 1 - (s_{i} - s_{j})\right)
\end{equation}
where $s_{i}$ and $s_{j}$ are the raw logits. The loss pushes the model to give higher logits to the positive examples than the negative ones, at least by a margin $m=1$. This loss is a differentiable surrogate for the AUC and has been demonstrated to improve class separation under severe imbalance (26\% positives in the dataset). If there is a single class in a batch, the loss is set to 0 to prevent degenerate gradients. Other training components (optimizer, scheduler, early stopping) remain unchanged.

\subsection{Baseline models}

The performance of the proposed ImprovedTCN is assessed against five other baseline models.

\begin{itemize}
\item \textbf{ImprovedTCN\_GRU:} A multi-scale TCN followed by a single-layer GRU with hidden size 108. The GRU processes the TCN outputs across time, and the final hidden state is classified.
\item \textbf{LSTM:} A two-layer Long Short-term Memory network and a dropout layer are implemented as a standard recurrent benchmark.
\item \textbf{TCN:} A vanilla TCN consisting of three stacked TCNBlock units. Each block uses a single dilation, kernel size, dropout, and residual connections. The output is mean-pooled over time and then classified.
\item \textbf{XGBoost:} Each 60-day sequence is flattened into a 1D vector. The XGBoost classifier is trained with 187 trees, maximum depth 7, and a binary logistic objective. No sequence structure is preserved.
\item \textbf{Random Forest:} Using the same flattened sequences, a Random Forest classifier is trained with 176 trees and a maximum depth of 14.
\end{itemize}

Optuna framework is used to find optimal hyperparameters for all models. All models are implemented in PyTorch and trained on an NVIDIA GPU. Hyperparameters are summarized in \autoref{tab:hyperparams}.

\begin{table}[htbp]
\centering
\footnotesize
\caption{Model Hyperparameters}
\label{tab:hyperparams}
\begin{tabular}{p{0.25\linewidth}p{0.35\linewidth}p{0.35\linewidth}}
\toprule
Model & Architecture & Key Hyperparameters \\
\midrule
ImprovedTCN (proposed) & 3 InceptionTCNBlocks + channel attention + pairwise ranking loss & num\_channels=50, dilations=[1,2,4], dropout=0.338, margin=1.0 (ranking loss), lr=0.000197 \\
ImprovedTCN\_GRU & 3 InceptionTCNBlocks + GRU & TCN channels=59, dilations=[1,2,4], GRU hidden=108, layers=1, dropout=0.241, lr=0.0095 \\
LSTM & 2 LSTM layers & hidden=70, dropout=0.201, lr=0.00066 \\
TCN & 3 TCN blocks & Dilations=[1,2,4], kernel=3, channels=91, dropout=0.261, lr=0.00449 \\
XGBoost & Flattened sequences & n\_estimators=187, max\_depth=7, lr=0.2595 \\
Random Forest & Flattened sequences & n\_estimators=176, max\_depth=14 \\
\bottomrule
\end{tabular}
\end{table}

\subsection{Profit-based Threshold Optimization}

Standard binary classification uses a fixed threshold of 0.5 to convert predicted probabilities into class labels. In a trading scenario, however, this is skewed because the costs of false positives (a losing trade) far outweigh the benefits of true positives (a winning trade). The decision rule is optimized directly on a profit function to align with the trader's economic goal.

The profit function is defined as follows:
\begin{equation}
\text{Profit} = TP \times g - (TP + FP) \times c
\end{equation}
where:
\begin{itemize}
\item TP (true positives) = number of correct $>5\%$ return predictions,
\item FP (false positives) = number of incorrect predictions,
\item $g = 0.05$ (5\% gross gain per correct trade),
\item $c = 0.001$ (0.1\% round-trip transaction cost per trade).
\end{itemize}

\subsection{Evaluation Metrics}

With severe class imbalance between positive and negative directional movements of Bitcoin (26.04\% positive events for the threshold of $>5\%$ return over 7 days), traditional classification measures such as accuracy are not suitable for model comparison and selection. In addition, this study goes beyond classification performance and also considers real trading profitability. Hence, this section introduces a dual-metric evaluation framework: (i) AUC for class separation and (ii) a profit-optimized decision threshold aligned with financial goals.

\subsubsection{Area under ROC curve (AUC)}

In this study, the Area Under the Receiver Operating Characteristic Curve is the primary assessment criterion. The AUC measures the model's ability to discriminate between positive and negative classes over all possible classification thresholds without specifying a decision boundary. Formally, the AUC is the same as the probability that a randomly selected positive instance will be scored higher than a randomly selected negative instance:
\begin{equation}
\text{AUC} = P(S_{p} > S_{n})
\end{equation}
where $S_{p}$ and $S_{n}$ are the predicted scores (logits or probabilities) of positive and negative examples, respectively. The empirical estimator of the AUC is given by the Mann-Whitney U statistic:
\begin{equation}
\text{AUC} = \frac{1}{|P||N|}\sum_{i \in P}\sum_{j \in N} \mathbb{I}(S_{i} > S_{j})
\end{equation}
The ties are managed by assigning a value of 0.5 to $\mathbb{I}(S_{i} > S_{j})$. This formulation highlights that AUC is a rank-based measure that is not sensitive to any particular classification threshold --- an inherent advantage given the challenge of class imbalance.

There are three key reasons AUC is the preferred metric:

\begin{itemize}
\item The AUC is not sensitive to class prior distribution, while accuracy can give an overly optimistic score by simply making the majority prediction, AUC does not. The present dataset is highly imbalanced (only 26\% belong to class 1 (7-day return $> 5\%$)), and the accuracy is a misleading metric. Unlike accuracy, which depends on absolute classification at a fixed threshold, AUC is based on relative prediction rankings and is more suitable for imbalanced financial time-series data.
\item Threshold-Independent Evaluation: Most financial forecasting models are not applied with a fixed cutoff of 0.5. Rather, traders fine-tune the threshold according to their risk appetite, transaction costs, and capital constraints. The AUC measures the quality of the model without relying on any cost-sensitive threshold optimization and therefore gives a threshold-free idea about the discriminative power of the model.
\item The proposed model is trained directly with a pairwise ranking loss, which is set to maximize the AUC score. Unlike accuracy or cross-entropy loss, AUC evaluates the model's performance in a manner that is consistent with its optimization objective.
\end{itemize}

\subsubsection{Profit-Based Threshold Optimization}

AUC is a good indicator of discriminative performance but is not directly convertible to economic value. In the real trading environment, false positives (incorrect predictions of $>5\%$ moves) incur transaction costs, while false negatives (missed $>5\%$ moves) represent lost profit opportunities. Thus, the conventional 0.5 probability threshold may not be optimal for decision-making. As defined in Section 3.6, this study uses a profit-optimized decision threshold that directly optimizes a trading profit function.

The profit function reflects the economically relevant trade-off: a 5\% gross return to each correct trade and a 0.1\% transaction cost per trade (regardless of correctness). The direct cost of false negatives is not obvious, but they represent missed opportunities. The optimal threshold $\tau^{*}$ is found via a grid search algorithm, and the choice of threshold that maximizes the profit function on the validation set is selected:
\begin{equation}
\tau^{*} = \arg\max_{\tau} \left[ TP(\tau) \times 0.05 - \left( TP(\tau) + FP(\tau) \right) \times 0.001 \right]
\end{equation}

An advantage of this method is that it overcomes the drawback of the traditional classification evaluation approach, where all errors are assumed to be equally costly. The cost structure is asymmetric in quantitative trading, and the optimal decision rule should reflect this asymmetry.

The profit-based threshold has two important purposes:

\begin{itemize}
\item Models are evaluated by the profit they generate under optimal threshold selection, making the model selection process less statistical and more aligned with financial goals.
\item Practical Decision Support: The threshold can be used directly in a live trading system as a clear decision rule: When the predicted probability exceeds the threshold, execute a trade at $\tau^{*}$. This bridges the gap between the probabilistic forecast and discrete trading actions.
\end{itemize}

\subsubsection{Secondary Metrics}

Even though AUC and profit-based performance are the primary metrics, the following secondary metrics are also reported to give a comprehensive overview:

\begin{itemize}
\item Precision: $\frac{TP}{(TP + FP)}$ -- the proportion of trades that are profitable
\item Recall: $\frac{TP}{(TP + FN)}$ - the proportion of actual positive moves that are captured
\item F1-Score: The harmonic mean of precision and recall
\item Confusion Matrix: Tabulated counts of TP, FP, TN, FN at the optimal profit threshold
\end{itemize}

These metrics are threshold-dependent and should be interpreted in relation to the profit function. Accuracy is not reported, as it has been shown to be invalid for highly imbalanced financial data.

\section{Results}

In this section, the results of the proposed model are empirically compared with five baseline models. All results stem from a rigorous walk-forward validation protocol that includes five growing training windows, and each is tested on an out-of-sample test window of 100 days. Two complementary goals are emphasized: (i) discriminative performance, measured by AUC, and (ii) economic performance, expressed as a profit-optimized decision threshold that explicitly accounts for transaction costs.

\subsection{Model Comparison: AUC Performance}

The aggregate AUC scores for all six models are presented in \autoref{tab:auc}, computed on the concatenated predicted probabilities across all five test folds. The AUC is the key indicator because it is highly robust to the extreme data imbalance.

\begin{table}[htbp]
\centering
\caption{AUC comparison across all models}
\label{tab:auc}
\begin{tabular}{lc}
\toprule
Model & AUC \\
\midrule
\textbf{ImprovedTCN (proposed)} & \textbf{0.6316} \\
TCN & 0.5692 \\
ImprovedTCN\_GRU & 0.5314 \\
XGBoost & 0.5011 \\
LSTM & 0.4586 \\
Random Forest & 0.3627 \\
\bottomrule
\end{tabular}
\end{table}

The proposed ImprovedTCN with an AUC value of 0.6316 outperforms the other five baseline models. The standard TCN model reached 0.5692, while the hybrid ImprovedTCN\_GRU gained 0.5316. Tree-based models and the LSTM model performed weaker, with XGBoost at 0.5011, LSTM at 0.4586, and Random Forest at 0.3627, which showed the least discriminative power.

\begin{figure}[htbp]
\centering
\includegraphics[width=\linewidth]{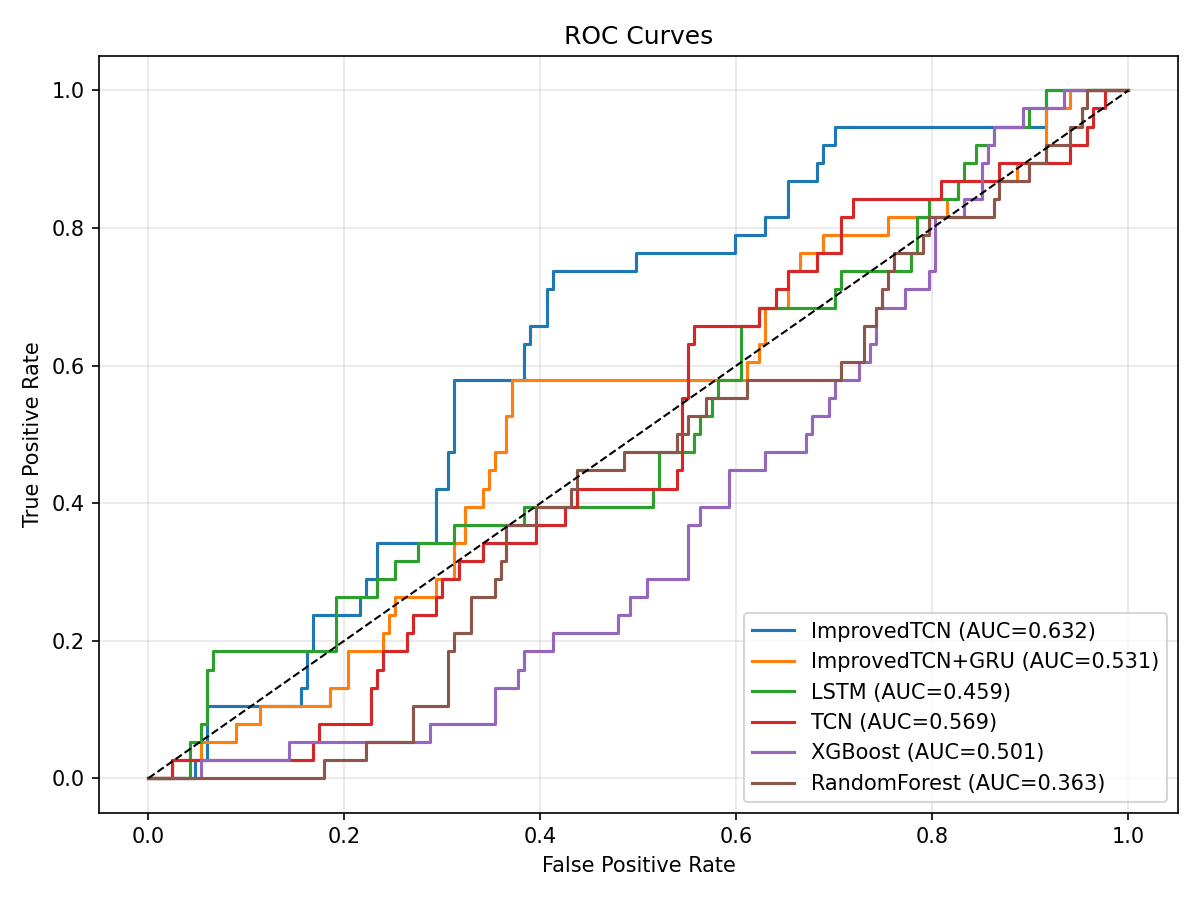}
\caption{ROC curves for all six models; the ImprovedTCN curve lies consistently above the others, visually confirming the AUC ranking.}
\label{fig:roc}
\end{figure}

To better illustrate the model's performance, \autoref{fig:roc} is presented. The ImprovedTCN curve is above all other curves, indicating a higher True Positive Rate and a lower False Positive Rate compared with other models. The pairwise ranking loss helped to improve the calibration and ranking quality in the proposed architecture.

\subsection{Profit-Optimized Threshold Results}

Even though AUC is a powerful indicator of discrimination, it does not directly translate into economic value. The difference between metrics is bridged by tuning the decision threshold to maximize the profit function for each model:
\begin{equation}
\text{Profit}(\tau) = TP(\tau) \times 0.05 - \left( TP(\tau) + FP(\tau) \right) \times 0.001
\end{equation}
where the threshold $\tau$ is swept over the interval $[0.01,0.99]$. \autoref{tab:profit} reports the optimal threshold, the associated profit, and confusion matrix counts for each model.

The proposed ImprovedTCN achieves the highest simulated profit (+1.703), compared with the hybrid GRU variant (+1.657) and substantially surpasses the tree-based models. Most importantly, the optimal threshold for ImprovedTCN is 0.11, which is significantly higher than the threshold that was chosen for ImprovedTCN\_GRU, LSTM, TCN, and XGBoost. A higher threshold requires more evidence before a trade is initiated. The model will only send a signal if the probability is more than 11\%, effectively ignoring low-confidence predictions and reducing false signals. While ImprovedTCN generates 159 signals (as compared to 137 for LSTM and 173 for TCN), the key economic advantage lies in the quality of these signals rather than their quantity.

\begin{table}[htbp]
\centering
\footnotesize
\caption{Profit optimized threshold and trading performance.}
\label{tab:profit}
\begin{tabular}{lccccccccc}
\toprule
\textbf{Model} & \textbf{Optimal $\tau$} & \textbf{Profit} & \textbf{TP} & \textbf{FP} & \textbf{FN} & \textbf{TN} & \textbf{Precision (\%)} & \textbf{F1} & \textbf{Recall (\%)} \\
\midrule
\textbf{ImprovedTCN} & \textbf{0.11} & \textbf{1.703} &  \textbf{38} &  \textbf{159} &  \textbf{0} &  \textbf{8}  &  \textbf{19.3} &  \textbf{0.3234} &  \textbf{100.0 }\\
ImprovedTCN\_GRU & 0.02 & 1.657 & 37 & 156 & 1 & 11 & 19.2 & 0.3203 & 97.4 \\
LSTM & 0.05 & 1.284 & 29 & 137 & 3 & 36 & 17.5 & 0.2929 & 90.6 \\
TCN & 0.05 & 1.395 & 32 & 173 & 0 & 0  & 15.6 & 0.2700 & 100.0 \\
XGBoost & 0.02 & 0.740 & 17 & 93 & 15 & 80 & 15.5 & 0.2394 & 53.1 \\
Random Forest & 0.22 & 1.271 & 29 & 150 & 3 & 23 & 16.2 & 0.2749 & 90.6 \\
\bottomrule
\end{tabular}
\end{table}

\subsection{Confusion Matrix}

\begin{figure}[htbp]
\centering
\includegraphics[width=\linewidth]{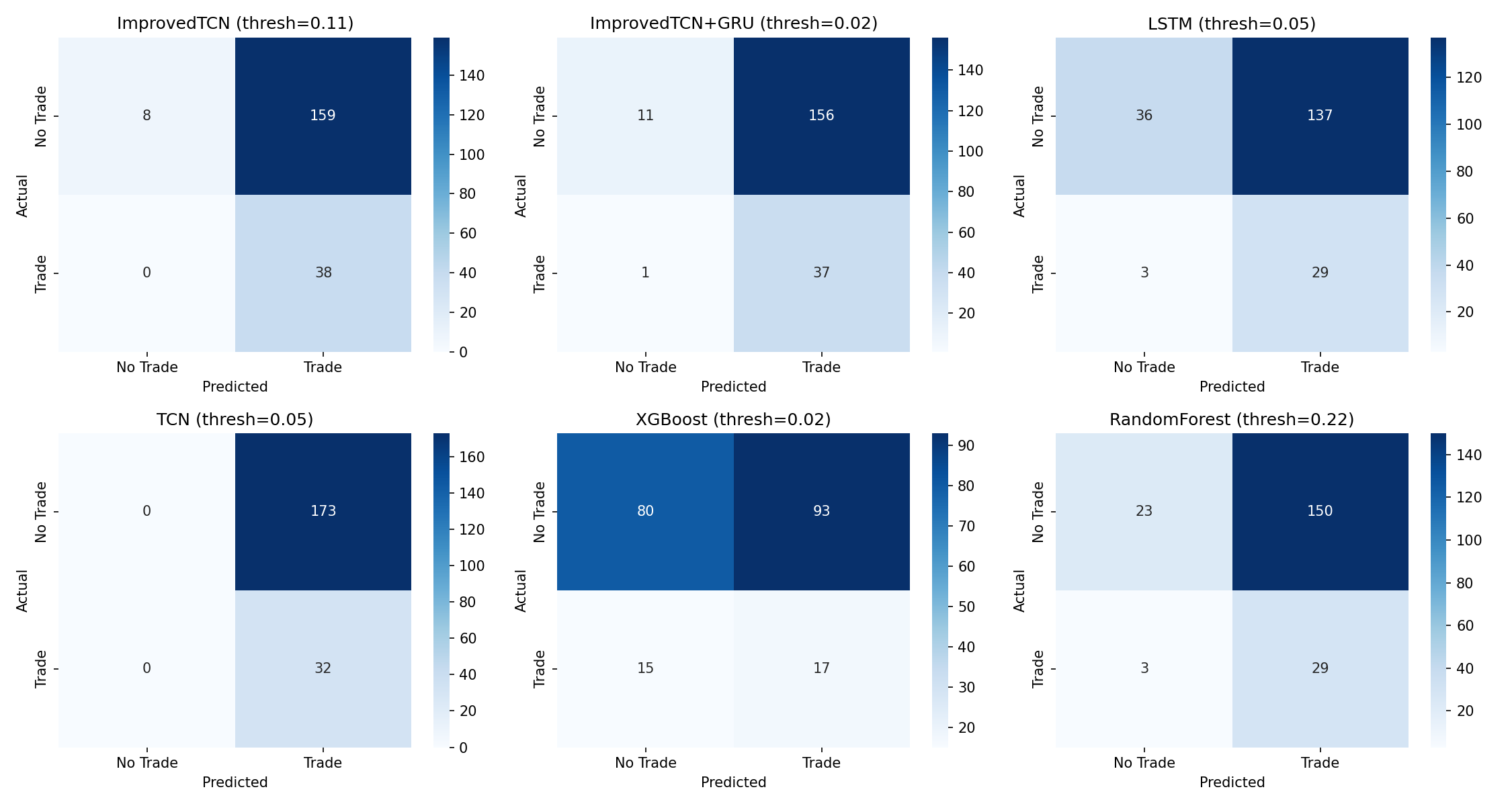}
\caption{Confusion matrices for all six models at optimal thresholds. ImprovedTCN matrix shows higher TN and lower FP compared to baselines.}
\label{fig:confusion}
\end{figure}

The confusion matrices for all models are shown in \autoref{fig:confusion}. The tabulated results reveal several observations.

\begin{itemize}
\item Complete Recall (100\%): The ImprovedTCN captures 38 positive events (returns $>5\%$ over 7 days); the ImprovedTCN\_GRU, TCN, LSTM, and Random Forest capture 37, 32, 29, and 29 positive events, respectively. This is a complete recall as a direct result of the optimization procedure for profit, which seeks to minimize the opportunity cost of false negatives over false positives.
\item The proposed ImprovedTCN achieves the highest precision (19.3\%) of all models. While this is not a large absolute number, it is an improvement compared to other models. Even small improvements in precision can make a major difference in risk-adjusted returns in a low-signal/high-noise financial environment.
\item LSTM correctly identifies 36 true negatives, but at the expense of reduced recall (90.6\%) -- it has missed 9 profitable opportunities compared to ImprovedTCN. The TCN model has zero false negatives because its threshold is so low that it catches every single positive event, but it does so by classifying everything as positive.
\end{itemize}

\section{Discussion}

The empirical results show that the proposed ImprovedTCN model outperforms all five baseline models, as its architectural, training, and algorithmic design considerations are particularly well-suited to the characteristics of financial time-series data.

The multi-scale, attention-weighted temporal framework is primarily responsible for the ImprovedTCN outperforming the vanilla TCN by an AUC of 0.6316 compared with 0.5692. Bitcoin's daily price actions are not linear, but rather involve multi-frequency feedback systems; microstructural liquidity alternates with medium-term sentiment cycles and long-term on-chain cycles of accumulation. In a standard TCN, all the feature maps are dilated with the same parameters, and it is assumed that the temporal velocity is the same in all feature maps. The InceptionTCN blocks, on the other hand, have dilated convolutional paths running in parallel. This enables the architecture to simultaneously manage short-term shocks (dilation = 1) and structural changes of the macro world (dilation = 4), while not increasing the parameter space too much.

Moreover, the standard deep learning models process input variables in an identical manner and are extremely sensitive to multicollinearity. Empirical correlation analysis (\autoref{tab:correlation} and \autoref{fig:corr_matrix}) shows that the fundamental on-chain parameters are highly correlated with each other. A standard LSTM or deep neural network will overfit and suffer from gradient instability when fed these raw features because they contain redundant signals as well as orthogonal predictive signals such as the Fear \& Greed Index or trading volume. The CNN channel attention mechanism calculates dynamic, real-time importance weights across channels by performing global average pooling over the temporal dimension of the data, in a context-dependent manner. The model can automatically down-weight highly correlated variables on the chain, while amplifying orthogonal sentiment signals, as the market enters speculative, high-volatility periods.

The suboptimal out-of-sample performance of the standard LSTM (AUC = 0.4586) demonstrates the well-known shortcomings of recurrent neural networks for financial forecasting. LSTMs have to receive sequences one step at a time; thus, the gradients have to pass through 60 steps of history sequentially. Such a long untruncated path of backpropagation can cause the network to suffer from exploding or vanishing gradients when faced with the heavy tails and high kurtosis of Bitcoin's returns (\autoref{tab:market_stats}). Moreover, the step-by-step memory update in an LSTM has difficulties with non-stationary financial time series because the past regime can be easily replaced by current market noise, which results in poor generalization ability out of sample. Structural data flattening is the reason behind the complete failure of the tree-based architectures: Random Forest (AUC = 0.3627). The 60-day temporal matrices need to be converted to large 1D arrays to process data using tree-based algorithms. The entire geometry, local dependencies, and the sequence of the features are completely lost. Without an inductive bias for temporal translation invariance, tree classifiers memorize the past combinations of inputs and fail to adapt when market trends are out of sample. However, XGBoost has the advantage of detecting the highest True Negatives among the other models, which demonstrates its capability in identifying spurious 5\% moves. Furthermore, introducing a GRU layer immediately after the multi-scale block in the hybrid ImprovedTCN\_GRU model decreased the AUC from 0.6316 to 0.5314. The performance drop shows that adding sequential bottlenecks to a well-calibrated and fully convolutional network creates optimization friction, reducing the gradient benefits of the underlying TCN structure. The economic aspect is that there is a fundamental misalignment between training a predictive model with a standard Binary Cross-Entropy (BCE) loss function and the trading goals. BCE minimizes overall logarithmic distance, which is a penalty on all errors equally distributed over the entire probability distribution. This pushes the model to make very conservative predictions that are not representative of the tail returns in highly imbalanced markets. The proposed architecture thus substantially changes the optimization objective to directly maximize the Area Under the ROC Curve, rather than BCE. The optimization procedure is compelled to try to assign a lower probability score to a negative instance than a positive instance. This model selection framework explicitly captures the asymmetric cost structure of quantitative trading in combination with a profit-maximizing decision threshold.

\section{Conclusion}

This study presents the ImprovedTCN, which is a multi-scale temporal convolutional network with Inception blocks, CNN channel attention, adaptive pooling, and pairwise ranking loss to predict whether Bitcoin price will exceed 5\% within the next 7 days. The model outperforms five strong baseline models with a higher AUC of 0.6316 and a higher profit of 1.703, using a rich multi-modal dataset (February 2018 - December 2025) and rigorous walk-forward validation.

Key contributions: (1) a parameter-efficient multi-scale TCN architecture for financial time series; (2) the use of channel attention and pairwise ranking loss for imbalanced, noisy data; (3) the explicit profit-threshold alignment between prediction and trading; and (4) extensive benchmarking that highlights the limitations of recurrent and tree-based approaches in this domain. In practice, the model provides a solid indicator for large Bitcoin price changes for quantitative traders and risk managers, and explicitly accounts for transaction fees. The positive profit factor provides a margin of safety for live deployment.

Further research is required to build upon this work for multi-asset portfolios, include uncertainty quantification, test the robustness of these policies to higher-frequency data, and develop hybrid reinforcement-learning policies for dynamic position sizing. The findings showcase the potential of deep-learning innovations with a focus on particular applications in generating economically relevant forecasts in volatile cryptocurrency markets, enriching the fields of machine learning and quantitative finance.


\end{document}